\documentclass[12pt,a4paper]{article}

\usepackage[left=2.5cm,top=2.5cm,bottom=2.5cm,right=2.5cm]{geometry}
\usepackage{graphicx}
\usepackage{amssymb}
\usepackage{amsthm}
\usepackage{amsmath}
\usepackage{amsfonts}
\usepackage{algorithm, algorithmic}
\usepackage{color}

\usepackage{cite}

\usepackage{enumitem}

\newtheorem{thm}{Theorem}[section]
\newtheorem{cor}[thm]{Corollary}
\newtheorem{lem}[thm]{Lemma}

\theoremstyle{definition}

\theoremstyle{remark}
\newtheorem{rem}{Remark}[section]

\numberwithin{equation}{section}

\begin{document}

\title{In-place associative integer sorting}


\author{A. Emre CETIN \\
email: aemre.cetin@gmail.com}

\maketitle

\begin{abstract}

A novel integer value-sorting technique is proposed replacing bucket sort, distribution counting sort and address calculation sort family of algorithms. It requires only constant amount of additional memory. The technique is inspired from one of the ordinal theories of ``serial order in behavior" and explained by the analogy with the three main stages in the formation and retrieval of memory in cognitive neuroscience namely (i) {\em practicing}, (ii) {\em storing} and (iii) {\em retrieval}. 

Although not suitable for integer rank-sorting where the problem is to put an array of elements into ascending or descending order by their numeric keys, each of which is an integer, the technique seems to be efficient and applicable to rank-sorting, as well as other problems such as hashing, searching, element distinction, succinct data structures, gaining space, etc. 

\end{abstract}


\section{Introduction}\label{sec:intro}

An integer value-sorting algorithm puts an array of {\em integers} into ascending or descending order by their {\em values}, whereas a rank-sorting algorithm puts an array of {\em elements} (satellite information) into ascending or descending order by their numeric {\em keys}, each of which is an integer. It is possible that a rank-sorting algorithm can be used in place of a value-sorting algorithm, since if each element of the array to be sorted is itself an integer and used as the key, then rank-sorting degenerates to value-sorting, but the converse is not always true.


The technique described in this study is suitable for arrays of integers where the integers are laid out in contiguous locations of the memory. Zero-based indexing is considered while accessing the integers, e.g., $A[0]$ and $A[n-1]$ are the first and last integers of the array, respectively, where $n$ is the number of integers of the array.

Nervous system is considered to be closely related and described with the ``serial order in behavior" in cognitive neuroscience~\cite{Lashley,Lashley_1} with three basic theories which cover almost all {\em abstract data types} used in computer science. These are chaining theory, positional theory and ordinal theory~\cite{Henson}.

Chaining theory is the extension of stimulus-response (reflex chain) theory, where each response can become the stimulus for the next~\cite{Henson}. From an information processing perspective, comparison based sorting algorithms that sort the arrays by making a series of decisions relying on comparing keys can be classified under chaining theory. Each comparison becomes the stimulus for the next. Hence, keys themselves are associated with each other. Some important examples are quick sort~\cite{Hoare}, shell sort~\cite{Shell}, merge sort~\cite{Burnetas} and heap sort~\cite{Williams}. 

Positional theory assumes order is stored by associating each element with its position in the sequence. The order is retrieved by using each position to cue its associated element. This is the method by which conventional (Von  Neumann) computers store and retrieve order, through routines accessing separate addresses in memory~\cite{Henson}. Content-based sorting algorithms where decisions rely on the contents of the keys can be classified under this theory. Each key is associated with a position depending on its content. Some important examples are distribution counting sort~\cite{Seward,Feurzig}, address calculation sort~\cite{Isaac,Tarter,Flores,Jones,Gupta,Suraweera}, bucket sort\cite{mahmoud:2000, Cormen} and radix sort~\cite{knuth:vol3,mahmoud:2000,sedgewick:algorithms_in_C, Cormen}.

Ordinal theory assumes order is stored along a single dimension, where that order is defined by relative rather than absolute values on that dimension. Order can be retrieved by moving along the dimension in one or the other direction. This theory need not assume either the item-item nor position-item associations of the previous theories~\cite{Henson}.

One of the ordinal theories of serial order in behavior is that of Shiffrin and Cook\cite{Shiffrin} which suggests a model for short-term forgetting of item and order information of the brain. It assumes associations between elements and a ``node'', but only the nodes are associated with one another. By moving inwards from nodes representing the start and end of the sequence, the associations between nodes allow the order of items to be reconstructed~\cite{Henson}.

The technique presented in this study is inspired from the ordinal model of Shiffrin and Cook. As in the ordinal model of Shiffrin and Cook, it is assumed that the associations are between the integers in the array space and the nodes in an imaginary linear subspace (ILS) that spans a predefined range of integers. The range of the integers spanned by the ILS is upper bounded by the number of integers $n$ but may be smaller and can be located anywhere provided that its boundaries do not cross over that of the array. This makes the technique in-place, i.e., beside the input array, only a constant amount of memory locations are used for storing counters and indices. An association between an integer in the array space and the ILS is created by a node using a monotone bijective hash function that maps the integers in the predefined interval to the ILS. When a particular distinct integer is mapped to the ILS, a node is created reserving all the bits of the integer except for the most significant bit (MSB) which is used to tag the word as a node of the ILS for interrogation purposes. The reserved bits become the record of  the node which then be used to count (practice) other occurrences of that particular integer that created the node. When all the key of the predefined interval are practiced, the nodes can be stored at the beginning of the array (short-term memory) retaining their relative order together with the information (cue) required to construct the sorted permutation of the practiced interval. Afterwards, the short-term memory is processed and the sorted permutation of the practiced interval is retrieved over the array space in linear time using only constant amount of additional memory. 

The adjective ``associative'' derived from two facts where the first one will be realized with the description of the technique in the following paragraphs. The second one is that, although it replaces all derivatives of the content based sorting algorithms such as distribution counting sort~\cite{Seward,Feurzig}, address calculation sort~\cite{Isaac,Tarter,Flores,Jones,Gupta,Suraweera} and bucket sort~\cite{mahmoud:2000,Cormen} on a RAM, it seems to be more efficient on a``content addressable memory'' (CAM) known as ``associative memory'' which in one word time find a matching segment in tag portion of the word and reaches the remainder of the word~\cite{Hanlon}. In the current version of associative sort developed on a RAM, the nodes of the imaginary linear subspace (tagged words) and the integers of the array space (untagged words) are processed sequentially which will be a matter of one word time for a CAM to retrieve previous or next tagged or untagged word.


\subsection{Definitions}\label{sec:pre}
Given an {\em array} $A$ of $n$ {\em integers}, $A[0], A[1],\ldots , A[n-1]$, the problem is to sort the integers in ascending or descending order. The notations used throughout the study are: 
\begin{enumerate} [label=({\roman{*}}), nosep]
\item Universe of integers is assumed $\mathbb{U} = [ 0 \ldots 2^{w}-1]$ where $w$ is the fixed word length.

\item Maximum and minimum integers of an array are, $\max (A) = \max(a \vert a \in A)$ and $\min (A) = \min(a \vert a \in A)$, respectively. Hence, range of the integers is, $m = \max (A) - \min (A) + 1$.

\item The notation $B \subset A$ is used to indicated that $B$ is a proper subset of $A$.

\item For two arrays $A_{1}$ and $A_{2}$, $\max (A_{1}) < \min (A_{2})$ implies $A_{1} < A_{2}$.

\end{enumerate}

\section{In-place Associative Integer Sorting}

The main difficulties of all distributive sorting algorithms is that, when the keys are distributed using a hash function according to their content, several of them may be clustered around a loci, and several may be mapped to the same location. These problems are solved by inherent three basic steps of associative sort (i) {\em practicing}, (ii) {\em storing} and (iii) {\em retrieval} which are the three main stages in the formation and retrieval of memory in cognitive neuroscience:
\begin{enumerate}[label=(\roman{*}), nosep,leftmargin=0pt, itemindent=* , nosep] 
\item Encoding or registration: receiving, processing and combining of received information.
\item Storage: creation of a permanent record of the encoded information.
\item Retrieval, recall or recollection: calling back the stored information in response to some cue for use in a process or activity.
\end{enumerate}

\subsection{Practicing}

An association between an integer in the array space and the ILS is created by a node using a monotone bijective hash function that maps the integers in the predefined interval to the ILS. The process of creating a node by mapping a distinct integer to the ILS is ``practicing a distinct integer of an interval''. Since ILS is defined on the array space, mapping a distinct integer to the ILS is just an exchange operation. Once a node is created, the redundancy due to the association between the integer and the position of the node (the position where the integer is mapped) releases the word allocated to the integer in the physical memory except for one bit which tags the word as a node for interrogation purposes. The tag bit discriminates the word as node and the position of the node lets the integer be retrieved back from the ILS using the inverse hash function.  This is ``integer retrieval". All the bits of the node except the tag bit can be cleared and used to store any information. Hence, they are the ``record'' of the node and the information stored in the record is the ``cue'' by which cognitive neuro-scientists describe the way that the brain recalls the successive items in an order during retrieval. For instance, it will be foreknown from the tag bit that a node has already been created while another occurrence of that particular integer is being practiced providing the opportunity to count other occurrences. The process of counting other occurrences of a particular integer is ``practicing all the integers of an interval'', i.e., rehearsing used by cognitive neuro-scientists to describe the way the brain manipulates the sequence before storing in a short (or long) term memory. 

Practicing does not need to alter the value of other occurrences. Only the first occurrence is altered while being practiced from where a node is created. All other occurrences of that particular integer remain in the array space but become meaningless. Hence they are ``idle integers''. On the other hand, practicing does not need to alter the position of idle integers as well, unless another distinct integer creates a node exactly at the position of an idle integer while being practiced. In such a case, the idle integer is moved to the former position of the integer that creates the new node. This makes associative sort unstable, i.e., equal integers may not retain their original relative order. However, an ILS can create other subspaces and associations using the idle integers that were already practiced by manipulating either their position or value or both. Hence, a part of linear algebra and related fields of mathematics can be applied on subspaces to solve such problems.

\begin{description}[leftmargin = 0pt]

\item[{\bf Universe of Integers.}] When an integer is first practiced, a node is created releasing $w$ bits of the integer free. One bit is used to tag the word as a node. Hence, it is reasonable to doubt that the tag bit limits the universe of integers because all the integers should be untagged and in the range $[0,2^{w-1}-1]$ before being practiced. But, we can,
\begin{enumerate}[label=(\roman{*}), itemindent = * , nosep]
\item partition $A$ into $2$ disjoint sublists $A_1 < 2^{w-1} \le A_2$ in $\mathcal{O}(n)$ time with well known in-place partitioning algorithms as well in a stable manner with~\cite{Katajainen},
\item shift all the integers of $A_2$ by $-2^{w-1}$, sort $A_1$ and $A_2$ associatively and shift $A_2$ by $2^{w-1}$.
\end{enumerate}
There are other methods to overcome this problem. For instance, 
\begin{enumerate}[label=(\roman{*}), itemindent = * , nosep]
\item sort the sublist $A[0\ldots (n/ \log n)-1]$ using the optimal in-place merge sort~\cite{Salowe},
\item compress $A[0\ldots (n/ \log n)-1]$ by Lemma~1 of~\cite{Franceschini_1} generating $\Omega(n)$ free bits,
\item sort $A[(n/ \log n)\ldots n-1]$ associatively using $\Omega(n)$ free bits as tag bits,
\item uncompress $A[0\ldots (n/ \log n)-1]$ and merge the two sorted sublists in-place in linear time by~\cite{Salowe}.
\end{enumerate}

\item [{\bf Number of Integers.}] If practicing a distinct integer lets us to use its $w-1$ bits to practice other occurrences of that particular integer, we have $w-1$ free bits by which we can count up to $2^{w-1}$ occurrences including the first integer that created the node. Hence, it is reasonable to doubt again that there is another restriction on the size of the arrays, i.e., $n \le 2^{w-1}$ under the assumption that an integer may always occur more than $2^{w-1}$ times for an array of $n > 2^{w-1}$. But an array can be divided into two parts in $\mathcal{O}(1)$ time and those parts can be merged in-place in linear time by~\cite{Salowe} after sorted associatively.

\end{description}

Hence, for the sake of simplicity, it will be assumed that $n \le 2^{w-1}$ and all the integers are in the range $[0,2^{w-1}-1]$ throughout the study.

\subsection{Storage}
Once all the integers in the predefined interval are practiced, the nodes dispersed in the ILS are clustered in a systematic way closing the distance between them to a direction retaining their relative order. This is the {\em storage} phase of associative sort where the received, processed and combined information required to construct the sorted permutation of the practiced interval is stored in the short-term memory (e.g., beginning of the array). When the nodes are moved towards a direction, it is not possible to retain the association between the ILS and array space. However, the record of a node can be further used to encode the absolute (former) position of that node as well, or maybe the relative position (with respect to the ILS) or how much that node is moved relative to its absolute or relative position during storing. Unfortunately, this requires that a record is enough to store both the positional information and the number of idle integers practiced by that node. However, as explained earlier, further associations can be created using the idle integers that were already practiced by manipulating either their position or value or both. Hence, if the record is enough, it can store both the positional information and the number of idle integers. If not, an idle integer can be associated accompanying the node to supply additional space to store the positional information. This definition immediately reminds one of the main difficulties of distributive sorting algorithms. When the keys are distributed using a hash function according to their content, several of them may be clustered around a loci. Hence, it is reasonable to think the difficulty in associating an idle integer accompanying the node. However, as explained earlier, the ILS can be defined anywhere on the array space and the range of the integers spanned by the ILS is upper bounded by $n$ but may be smaller and can be located anywhere making the technique in-place.

\subsection{Retrieval}

Finally, the sorted permutation of the practiced interval is constructed in the array space, using the information stored in the short-term memory. This is the {\em retrieval} phase of associative sort which depends on the information stored in the record of a particular node. If the record is enough, it stores both the position of the node and the number of practiced idle integers. If not, an associated idle integer accompanying the node stores the position of the node while the record holds the number of practiced idle integers. The positional information cues the recall of the integer using the inverse hash function. This is ``integer retrieval'' from imaginary subpace. Hence, the retrieved (recalled) integer can be copied on the array space as many as it occurrs. It should be noticed that one can process the information in the short-term memory from right to left and distinguish an idle key (untagged word) from a node (tagged word). From right to left, an (untagged) idle key implies that it is accompanying the preceding (tagged) node for additional storage. 

Hence, moving through nodes that represent the start and end of practiced integers as well as retaining their relative associations with each other even when their positions are altered by cuing allow the order of integers to be constructed in linear time in-place.

From complexity point of view, associative sort shows similar characteristics with bucket sort and distribution counting sort. Hence, it can be thought of as {\em in-place associative bucket sort} or {\em in-place associative distribution counting sort}. Distribution counting sort is seldom discussed in the literature although it has been around more than 50 years since proposed by Seward~\cite{Seward} in 1954 and by Feurzig \cite{Feurzig} in 1960, independently, and known to be the method that makes radix sort possible on digital computer. It is known to be very powerful when the integers have small range. Given $n$ integers $A[0\ldots n-1]$ each in the range $[0,m-1]$, its time-complexity is $\mathcal{O}(n+m)$ and requires $n+m$ additional space for a stable and $m$ for an unstable sort. Hence, distribution counting sort becomes efficient and practical when $m=\mathcal{O}(n)$ defining its time-space trade-offs. On the other hand, bucket sort is a generalization of distribution counting sort. In fact, if each bucket has size $1$, then bucket sort degenerates to distribution counting sort. However, the variable bucket size allows it to use $\mathcal{O}(n)$ memory instead of $\mathcal{O}(m+n)$ memory. Its average case time complexity is $\mathcal{O}(n+m)$ and if $m=\mathcal{O}(n)$, then it becomes $\mathcal{O}(n)$. Its worst case time complexity is $\mathcal{O}(n^2)$.

With this introductory information, the contribution of this study is,
\begin{description}[leftmargin=0pt]
\item[{\bf A practical sorting algorithm}] that sorts $n$ integers $A[0\ldots n-1]$ each in the range $[0,m-1]$ using $\mathcal{O}(1)$ extra space in $\mathcal{O}(n+m)$ time for the worst, $\mathcal{O}(m)$ time for the average (uniformly distributed integers) and $\mathcal{O}(n)$ time for the best case. The ratio $\frac{m}{n}$ defines the efficiency (time-space trade-offs) of the algorithm letting very large arrays to be sorted in-place. The algorithm is simple and practical replacing bucket sort, distribution counting sort and address calculation sort family of algorithms improving the space requirement to only $\mathcal{O}(1)$ extra words.

Practical comparisons for $1$ million 32 bit integers with quick sort showed that associative sort is roughly $2$ times faster for uniformly distributed integers when $m=n$. When $\frac{m}{n} = 10$ performances are same. When $\frac{m}{n} = \frac{1}{10}$ associative sort becomes more than $3$ times faster than quick sort. If the distribution is exponential, associative sort shows better performance up to $\frac{m}{n} \approx 25$ when compared with quick sort. 

Practical comparisons for $1$ million 32 bit integers showed that radix sort is $2$ times faster for uniformly distributed integers when $m=n$. However, associative sort is slightly better than radix sort when $\frac{m}{n} = \frac{1}{10}$. Further decreasing the ratio to $\frac{m}{n} =\frac{1}{100}$, associative sort becomes more than $2$ times faster than radix sort. 

Practical comparisons for $1$ million 32 bit integers showed that value-sorting version of distribution counting sort (frequency counting sort~\cite{mahmoud:2000}) is $2$ times faster than associative sort for $\frac{m}{n} = 1$. Distribution counting sort is still slightly better but the performances get closer when $\frac{m}{n} < \frac{1}{10}$ and $\frac{m}{n} > 10$.  

Even omitting its space efficiency for a moment, associative sort asymptotically outperforms all content based sorting algorithms when $n$ is large relative to $m$. 

\end{description}

The technique seems to be efficient and applicable for other problems, as well, such as hashing, searching, element distinction, succinct data structures, gaining space, etc. For instance, there are several space gaining techniques available and widely used in the literature for in-place and minimum space algorithms~\cite{Andersson_1,Franceschini_1,Katajainen,Katajainen_1}. However, as known to the author, all these in-place and minimum space algorithms have a dedicated explicit technique that is used only for space gaining purpose. On the contrary, gaining space is an inherent step of associative sort which improves its performance and can be used explicitly.



\section{Basics of Associative Sort} \label{sec:basics}

Given $n$ {\em distinct} integers $A[0\ldots n-1]$ each in the range $[\delta, \delta+m-1]$ where $\delta=\min(A)$, if $m=n$ and $A$ is the sorted permutation, then there is a bijective relation $i=A[i]-\delta$ between each integer and its position. From contradiction, $i \ne A[i]-\delta$ implies that the integer $A[i]$ is not at its exact position. Its exact position can be calculated by $j=A[i]-\delta$. Therefore, the simple monotone bijective hash function $j=A[i]-\delta$ that maps the integers to $j \in [0,n-1]$ can sort the array in $\mathcal{O}(n)$ time using $\mathcal{O}(1)$ constant space. This is in-situ permutation (cycle leader permutation)  where $A$ is re-arranged by following the cycles of a permutation $\pi$. First $A[0]$ is sent to its final position $\pi(0)$ (calculated by $j=A[i]-\delta$). Then the element that was in $\pi(0)$ is sent to its final position $\pi(\pi(0))$. The process proceeds in this way until the cycle is closed, that is until the integer addressing the first position is found which means that the association $0 = A[0] - \delta$ is constructed between the integer and its position. Then the iterator is increased to continue with $A[1]$. At the end, when all the cycles of $A[i]$ for $i=0,1..,n-1$ are processed, all the integers are in their exact position and the association $i = A[i] - \delta$ is constructed between the integers and their position resulting in the sorted permutation.

If we look at this approach closer, we can interpret the technique entirely different. That is, we are indeed creating an ILS $Im[0\ldots n-1]$ over $A[0\ldots n-1]$ where the relative basis of this ILS coincides with that of the array space in the physical memory. The ILS spans a predefined interval of the range of integers depending on $n$. Since $m=n$, it spans the entire range of integers. The association between the array space and the ILS is created by a node using the monotone bijective hash function $i = A[i] - \delta$ that maps a particular integer to the ILS. When a node is created for a particular integer, the redundancy due to the association between the integer and the position of the node releases the word allocated for the integer in the physical memory. Hence, we can clear the node ($A[i]=0)$ and set its tag bit, for instance its most significant bit (MSB) to discriminate it as a node, and use the remaining $w-1$ bits of the node for any other purpose. When we want the integer back to array space from ILS, we can use the inverse of hash function and get the integer back by $A[i]=i + \delta$ to the array space. However, we don't use free bits of a node for other purposes in this case because it is known that all the integers are distinct and hence only one integer will be practiced at a location creating a node. Therefore, instead of tagging the word as node using its MSB, we use the integer itself to tag the word ``implicitly'' as node, since if an integer is mapped to the ILS, then it will always satisfy the hash function $i = A[i] - \delta$. Hence, the integers are ``implicitly practiced'' in this case.

Mathematically, consider an array of $n$ {\em distinct} integers $A[0\ldots n-1]$ each in the range $[0,\mathbb{U}]$ and stored sequentially in the RAM. Let $\mathbb{U}$ denote the field of positive integers including $0$ and consider the elements of the array as a set of 3-tuples $\mathbf{x} = [i, \: A[i], \: 1]$ of integers forming a vector space over $\mathbb{U}$ denoted by $\mathbb{U}^3$. Hence, $A[0\ldots n-1]$ is a 3-dimensional vector space over $\mathbb{U}$ and any element of the array is represented by 3 integer components where the first one in $[0,n-1]$ represents the index, i.e., the position of the element in the array, the second one in $[0,\mathbb{U}]$ represents the element (either an integer or a node) stored at that position, and the third one is a dummy constant.

Now, consider an ILS $Im[0\ldots n-1]$ which is a vector space over a given interval of range of integers of $A$ as a set of 3-tuples $\mathbf{x} = [i', \: A[i'], \: 1]$ of integers. This time, the second component is a subset of integers of $A$ in the range $[a,b]$ with $b-a<n$. It should be noted that, for an array of $n$ integers each in the range $[\delta,\delta+m-1]$ where $\delta=\min(S)$, if $m=n$, i.e., the number of integers is equal to the range of integers, then the ILS spans the entire range of integers.

A bijective linear mapping from the array space to the ILS can be defined as,
\begin{equation}\label{eqn:transformation_1}
\begin{split}
\begin{bmatrix}
i' \\
A[i'] \\
1
\end{bmatrix}
=\begin{bmatrix}
0\quad & 1 & \quad -\delta \\
0\quad & 1 & \quad 0 \\
0\quad & 0 & \quad 1
\end{bmatrix}
\begin{bmatrix}
i \\
A[i] \\
1
\end{bmatrix}
=\begin{bmatrix}
A[i]-\delta \\
A[i] \\
1
\end{bmatrix}
\end{split}
\end{equation}
with $\delta = \min(A)$. Eqn.~\ref{eqn:transformation_1} tells us that when an integer is mapped into the ILS, its new position will be $i' = A[i] - \delta$ in both the ILS and the array space while its value is unchanged, i.e., $A[i'] = A[i]$. From algorithm point of view, this is equivalent to swapping $A[i]$ with $A[i']$. The linear  mapping in Eqn.\ref{eqn:transformation_1} does not have an inverse. This is expectable when we consider that after swapping $A[i]$ with $A[i']$ there is no way to know where $A[i']$ was before. However, when we look at the right side of this equation closer, we immediately see the redundancy between the new position of an integer ($A[i] - \delta$) and its value ($A[i]$) provided that $\delta$ is known. This redundancy is the fact that makes cycle leader permutation possible. Therefore, cycle leader permutation is a special case of associative sort.

This mathematical definition gives us other opportunities. For instance, we can define our transformation matrix as,
\begin{equation}
\begin{split}
\begin{bmatrix}
i' \\
A[i'] \\
1
\end{bmatrix}
=\begin{bmatrix}
0\quad & 1 & \quad -\delta \\
0\quad & 0 & \quad 0 \\
0\quad & 0 & \quad 1
\end{bmatrix}
\begin{bmatrix}
i \\
A[i] \\
1
\end{bmatrix}
=\begin{bmatrix}
A[i]-\delta \\
0 \\
1
\end{bmatrix}
\end{split}
\end{equation}
which says that $w$ bits of an integer that is mapped into the ILS at $i'$ can be cleared and used for any other purpose because it can be retrieved back to any location (for instance to $j$) by,
\begin{equation}
\begin{split}
\begin{bmatrix}
i' \\
A[i'] \\
1
\end{bmatrix}
=\begin{bmatrix}
0\quad & 0 \quad & j \\
1\quad & 0 \quad & \delta \\
0\quad & 0 \quad & 1
\end{bmatrix}
\begin{bmatrix}
A[i]-\delta \\
0 \\
1
\end{bmatrix}
=\begin{bmatrix}
j \\
A[i] \\
1
\end{bmatrix}
\end{split}
\end{equation}
provided that $\delta = \min(A)$ is known. 

\begin{description}[leftmargin = 0pt]

\item [{\bf Imaginary linear subspace}] can be defined anywhere on the array space $A[0\ldots n-1]$ provided that its boundaries does not cross over that of the array space. An ILS can be defined as $Im[a\ldots b]$ with $b-a<n$. Two supplementary definitions should be given:
\begin{enumerate}[label=(\roman{*}), nosep] 
\item The relative basis of the subspace over the array space. This is defined by the statement $Im[a\ldots b]$ over $A[u\ldots v]$ which strictly implies that $v-u = b-a$.
\item The interval of range of integers spanned by $Im[a\ldots b]$. But, it is immediate that $Im[a\ldots b]$ can span any subset $A_1 \subset A$ where each integer of $A_1$ is in $[\beta,\beta+b-a]$ with $\beta \in A$.  
\end{enumerate}

Furthermore, an ILS of a small interval that is created casually anywhere on the array space can be moved with all its nodes to a given direction (left or right) with respect to the array space provided that the hash function that associates the subspace and the array space is shifted as much as the subspace is shifted relative to the array space. This is ``subspace shifting''.

\end{description}

\begin{description}[leftmargin = 0pt]
\item [{\bf Node}] is an association (interconnection) between the ILS and the array space. It is created by mapping an integer from the array space into the ILS during practicing. A monotone bijective hash function is used for the mapping. The necessary requirements that a particular integer of the array space can be mapped into an ILS creating a node are,
\begin{enumerate}[label=(\roman{*}), nosep] 
\item The integer should be in the interval of the range of integers spanned by the ILS.
\item There should not be a node already created by another occurrence of that particular integer.
\end{enumerate}

\end{description}


\section{Sorting $n$ Integers}\label{subsec:es_multiple}

In this section, the associative sorting technique will be introduced which is based on the three basic steps: (i) practicing, (ii) storing and (iii) retrieval.  

Using $w-1$ bits (record) of a node released while a particular integer is being practiced, other occurrences of that particular integer can be counted. Unfortunately, we need $\log n$ bits of the record to encode the absolute position of the node during storing. Hence, it is reasonable to doubt that we can count up to $2^{w-1-\log n}$ including the first occurrence that created the node. Fortunately, this is not the case. We can count using all $w-1$ bits of the record, and while the nodes are being stored at the beginning of the array (short-term memory), we can get an idle integer immediately after a node that has practiced at least $2^{w-1-\log n}$ integers and write the absolute position of that node over the idle integer as the cue. In this case, the record of the node stores only the number of idle integers practiced by that node. But this time, we encounter another serious problem of all distributive sorting algorithms. Depending on the distribution of the integers, the nodes that are created during practicing are dispersed over the ILS. If the nodes are clustered to the beginning, how an idle integer can be inserted immediately after a particular node if there is another node immediately before and after that particular node during storing? The answer is in the pigeonhole principle. Pigeonhole principle says that,

\begin{cor}
Given $n$ integers $A[0\ldots n-1]$, the maximum number of distinct integers that may occur contemporary in $A$ at least $2^{w-1-\log n}$ times is,
\begin{equation}
\lceil \frac{n}{2^{w-1-\log n}} \rceil
\end{equation}
\end{cor}
Hence, if the size of the array is say $n = 2^{w-1}$, the maximum number of {\em distinct} integers that may occur contemporary in $A$ at least $1$ time is $n$. But the node itself represents the first occurrence which creates it. Therefore, 
\begin{cor}
The maximum number of nodes that each can practice at least $2^{w-1-\log n}$ integers and hence need an idle integer immediately after itself during storing is equal to, 
\begin{equation}\label{eqn:epsilon}
\epsilon = \lceil \frac{n/2}{2^{w-1-\log n}} \rceil
\end{equation}
and upper bounded by $n/2$. 
\end{cor}
This means that, 
\begin{cor}
If the integers are practiced to $Im[\epsilon, n-1]$ over $A[\epsilon, n-1]$ where $\epsilon$ is calculated by Eqn.\ref{eqn:epsilon}, then there will be $\epsilon$ integers at the beginning of the array either idle or out of the practiced interval which will prevent collisions while inserting idle integers immediately after the nodes that practiced at least $2^{w-1-\log n}$ integers. 
\end{cor}
Hence,
\begin{lem}\label{lem:sorting_seq_of_distinct}
Given $n<=2^{w-1}$ integers $A[0...n-1]$ each in the range $[0,2^{w-1}-1]$, all the integers in the range $[\delta,\delta+n-\epsilon-1]$ where $\delta=\min(A)$ can be sorted associatively at the beginning of the array in $\mathcal{O}(n)$ time using $\mathcal{O}(1)$ constant space.
\end{lem}

\begin{proof}
Given $n<=2^{w-1}$  distinct integers $A[0...n-1]$ each in the range $[0,2^{w-1}-1]$, it is not possible to construct a monotone bijective hash function that maps all the integers of the array into $j \in [0,n-1]$ without additional storage space~\cite{Belazzougui}. However, a monotone bijective hash function can be constructed as a partial function~\cite{rosen:discrete_math_handbook} that assigns each integer of $A_1 \subset A$ in the range $[\delta,\delta+n-\epsilon-1]$ with $\delta=\min(A)$ to exactly one element in $j \in [\epsilon,n-1]$. The partial monotone bijective hash function of this form is,
\begin{equation}\label{eqn:hash_func_mul}
\begin{split}
j=A[i]- \delta + \epsilon \quad \text{if} \quad A[i] - \delta + \epsilon < n
\end{split}
\end{equation}

With this definition, the proof has three basic steps of associative sort: 
\begin{enumerate}[label=(\roman{*})]
\item Practice all the integers of the interval $[\delta,\delta+n-\epsilon-1]$ into $Im[\epsilon \ldots n-1]$ over $A[\epsilon \ldots n-1]$.

\item Store the nodes at the beginning of the array (short term memory) in order. If at least $2^{w-1-\log n}$ idle integers are practiced by a particular node, find the nearest idle integer searching to the right and move it immediately after that node and write the absolute position of the node over the idle integer by modifying it. Otherwise, i.e., if less than $2^{w-1-\log n}$ idle integers are practiced by a particular node, encode the absolute position of the node into $\log n$ bits of its record where the remaining ${w-1-\log n}$ bits store the number of idle integers. 

\item Retrieve the encoded information from the short term memory processing the records backwards to construct the sorted permutation of the practiced interval. If MSB of a record is $1$, then it is a node and its record stores both the absolute position of that node and the number of idle integers practiced by that node. Otherwise, i.e., if MSB of a record is $0$, then it is indeed an idle integer brought immediately after a node that has practiced at least $2^{w-1-\log n}$ idle integers. Hence, read the absolute position of the node from the idle integer and decode the number of idle integers from the record of the node (predecessor of idle integer).

\end{enumerate}
\end{proof}

\subsection{Practicing Phase}\label{sec:counting_mul}

Practicing is the process of encoding the necessary information required to recollect the sorted permutation of the practiced, i.e., received, processed and combined interval. An individual iteration over the array can practice only the distinct integers disregarding other occurrences in the predefined interval of the ILS creating nodes exactly equal to the number of distinct integers in that interval. Such an iteration is ``practicing distinct integers of an interval''. On the other hand, an iteration can practice all the integers of the array that are in the predefined interval of the ILS. For instance, if the number of distinct integers in a given interval that create a node is $n_d$ and the number of total occurrences other than those particular distinct integers is $n_c$, then $n_d$ nodes are created practicing (counting) other $n_c$ integers that become idle, hence become meaningless. Such an iteration is ``practicing all the integers of an interval" and one can inquiry the existence and number of occurrences of a given value in that interval in $\mathcal{O}(1)$ time after practicing.

\begin{enumerate}[label=\bf{Algorithm \Alph{*}.}, ref=Algorithm \Alph{*}, leftmargin=0pt, itemindent=*, start=1] 
\item \label{algorithm:es_fgl_mul} Practice all the integers of the interval $[\delta,\delta+n-\epsilon-1]$ into the ILS $Im[\epsilon\ldots n-1]$ over $A[\epsilon\ldots n-1]$ using Eqn.~\ref{eqn:hash_func_mul}. It is assumed that the minimum of the array $\delta=\min(A)$ is known and $\epsilon$ is calculated by Eqn.~\ref{eqn:epsilon}.
\end{enumerate}

\begin{enumerate}[label=\bf{A\arabic{*}.}, ref=A\arabic{*}, itemindent=* , nosep]
\item initialize $i = 0$;\label{algo17:item0}
\item if $A[i]<\delta$, then $A[i]$ is an idle integer of an interval that has already been sorted. Hence, increase $i$ by one and repeat this step; \label{algo17:item1}
\item if MSB of $A[i]$ is $1$, then $A[i]$ is a node. Hence, increase $i$ by one and goto step \ref{algo17:item1};\label{algo17:item2}
\item if $A[i] - \delta + \epsilon \ge n$, then $A[i]$ is an integer that is out of the practiced interval. Hence, increase $n_d'$ by one that counts the number of integers out of the practiced interval, update $\delta'=min(\delta', A[i])$, increase $i$ by one and goto to step \ref{algo17:item1};\label{algo17:item3}
\item otherwise, $A[i]$ is an integer to be practiced. Hence, calculate $j = A[i] - \delta + \epsilon$;\label{algo17:item4}
\item if MSB of $A[j]$ is $0$, then $A[i]$ is the first integer that will create the node at $j$. Move $A[j]$ to $A[i]$, clear $A[j]$ and set its MSB to $1$ making it a node. If $j \le i$ increase $i$ by one. Increase $n_d$ by one that counts the number of distinct integers (nodes), and goto step \ref{algo17:item1}; \label{algo17:item5}
\item otherwise, $A[j]$ is a node that has already been created. Hence, clear MSB of $A[j]$, increase $A[j]$ by one (number of idle integers) and set its MSB back to $1$. Increase both $i$ and $n_c$ by one ($n_c$ counts the number of total idle integers over all distinct integers) and goto step \ref{algo17:item1};
\end{enumerate}

\begin{rem}
It should be noted that, all the three phases of associative integer sorting can be implemented using the underlying signed integer notation of computers. As long as the number of negative integers are one more than positive integers, $-1$ can be used to tag a word as node. Afterwards, the node can be decreased by one for each practiced idle key. In such a case, there is no need to struggle with bitwise operations. For instance, if we consider \ref{algorithm:es_fgl_mul}, it becomes,

\begin{enumerate}[label=\bf{A\arabic{*}'.}, ref=A\arabic{*}', itemindent=* , nosep]
\item initialize $i = 0$;\label{algo17:item0_}
\item if $A[i]<\delta$, increase $i$ by one and repeat this step; \label{algo17:item1_}
\item if $A[i] - \delta + \epsilon \ge n$, increase $n_d'$ by one, update $\delta'=min(\delta', A[i])$, increase $i$ by one and goto \ref{algo17:item1_};\label{algo17:item3_}
\item otherwise, $A[i]$ is an integer to be practiced. Hence, calculate $j = A[i] - \delta + \epsilon$;\label{algo17:item4_}
\item if $A[j]>=0$, move $A[j]$ to $A[i]$, and set $A[j]=-1$ making it a node. If $j \le i$ increase $i$ by one. Increase $n_d$ by one, and goto step \ref{algo17:item1_}; \label{algo17:item5_}
\item otherwise, decrease $A[j]$ by one. Increase both $i$ and $n_c$ by one and goto step \ref{algo17:item1_};
\end{enumerate}
\end{rem}

\begin{description}[leftmargin = 0pt]
\item[{\bf Associative Range Queries.}] Instead of practicing all the integers in an interval as in \ref{algorithm:es_fgl_mul}, one can practice only the distinct integers in an interval, for instance $[\delta, \delta+n/2-1]$ into $Im[n/2 \ldots n-1]$  over $A[n/2 \ldots n-1]$, writing each created node's position to the previously created node's record. Hence, an associative linked array is obtained in $\mathcal{O}(n)$ time that can answer range queries such as: ``Which are the integers in the range $[\delta, \delta+n/2-1]$?'' in $\mathcal{O}(n_d)$ time. Furthermore, if a secondary ILS $Im[0 \ldots n/2-1]$ is created which does not overlap with the primary one, then {\em all} the idle integers of the same interval can be practiced and mapped to the secondary ILS. This constructs a further association between the subspaces through matching node positions with respect to each subspace basis. As a result, while answering range queries using the primary subspace, the number of integers can be queried with $\mathcal{O}(1)$ secondary subspace access. When finished, all the integers mapped to the imaginary subspaces can be retrieved back and the queries can continue with another interval of interest.
\end{description}

\subsection{Storing Phase}\label{sec:partitioning_mul}

Storing is the process of creating permanent records in the short-term memory in a systematic and organized way where the received, processed and combined information during practicing is encoded into the records of the nodes to cue the recall of the necessary information that will be used to construct the sorted permutation of the practiced interval. 

Practicing creates $n_d$ nodes and $n_c$ idle integers. This means $n_d$ distinct integers of the practiced interval are mapped into the ILS creating nodes that are dispersed with relative order in $Im[\epsilon\ldots n-1]$ over $A[\epsilon\ldots n-1]$ depending on the statistical distribution of the integers. On the other hand, $n_c$ idle integers are distributed disorderly together with $n_d'$ integers out of the practiced interval in the array space.

In storing phase, the nodes are clustered in a systematic way, i.e., the gaps between the nodes of the ILS are closed to a direction  without altering their relative order with respect to each other. When the nodes are moved towards a direction, it is not possible to retain the association between the nodes and the integers. Furthermore, we cannot freely use $\log n$ bits of a record of the node to encode its absolute position because the node has practiced other occurrences (idle integers) and if the number of the idle integers occupies more than ${w-1-\log n}$ bits, it would not be possible to encode the absolute position of the node into the record. But, the maximum number of nodes that will need an idle integer immediately after itself during storing is equal to, 
\begin{equation}
\epsilon = \lceil \frac{n/2}{2^{w-1-\log n}} \rceil
\end{equation}
Hence, by mapping the integers to $Im[\epsilon\ldots n-1]$ over $A[\epsilon\ldots n-1]$, we create a recovery area exactly equal to $\epsilon$ which prevents collisions during storing and lets us to bring an idle integer immediately after a particular node that has practiced at least $2^{w-1-\log n}$ idle integers.

\begin{enumerate}[label=\bf{Algorithm \Alph{*}.}, ref=Algorithm \Alph{*}, leftmargin=0pt, itemindent=*, start=2] 
\item \label{algorithm:es_fgp_mul} Store the encoded information of the practiced interval in the short term memory. If at least $2^{w-1-\log n}$ idle integers are practiced by a particular node, find the nearest idle integer on the right side searching forward  and move it immediately after that particular node and write the absolute position of the node over the idle integer by modifying it. Hence, the absolute position of the node can be recalled from the idle integer during retrieval. In such a case, the record of the node preceding the idle integer in the short term memory only stores the number of idle integers. Otherwise, i.e., if less than $2^{w-1-\log n}$ idle integers are practiced by a particular node, encode the absolute position of the node into $\log n$ free bits of the record together with the bits (other than the tag bit) that keep the number of idle integers. 
\end{enumerate}
\begin{enumerate}[label=\bf{B\arabic{*}.}, ref=B\arabic{*}, itemindent=* , nosep]
\item initialize $i = \epsilon$, $j = 0$, $k = n_d$, and $\epsilon' = 0$ which will count the exact number of nodes that have practiced at least $2^{w-1-\log n}$ idle integers;
\item if MSB of $A[i]$ is $0$, then $A[i]$ is either an idle integer or an integer that is out of the practiced interval. Hence, increase $i$ and repeat this step; \label{algo18:item1}
\item otherwise, $A[i]$ is a node. Hence, get the number of practiced idle integers into $s$; \label{algo18:item2}
\item if $s+1\le 2^{w-1-\log n}$, then encode the absolute position $i$ ($\log n$ bits) of the node into $s$, move $A[j]$ to $A[i]$, and write $s$ to $A[j]$. Increase $i$ and $j$ and decrease $k$. If $k = 0$ exit, otherwise goto step \ref{algo18:item1}; \label{algo18:item3}
\item otherwise, i.e., if $s+1>2^{w-1-\log n}$, then  swap $A[i]$ with $A[j]$. Find one of $n_c$ idle integers on the right starting a search from $A[p]$ where $p$ is either equal to $j+1$ if this is the first time that a search is started or equal to the last idle integer position found in the previous search. Move $A[j+1]$ to this safe location ($A[p]$) and write the absolute position $i$ of the node over $A[j+1]$. Increase $i$ and $\epsilon'$ by one and $j$ by two and decrease $k$ by one. If $k = 0$ exit, otherwise goto step \ref{algo18:item1};\label{algo18:item6} \label{algo18:item4}
\end{enumerate}


\subsection{Retrieval Phase}\label{sec:decoding_mul}

Retrieval is the reverse of storing. The sorted permutation of the practiced interval is constructed using the stored information in the short-term memory. On the other hand, getting the integer from the ILS back to the array space is integer retrieval where the positional information stored in the record of a node cues the recall of the integer using the inverse hash function.

Storing clusters $n_d$ nodes and $\epsilon'$ idle integers at $A[0 \ldots n_d+\epsilon'-1]$  with the necessary information required to construct the sorted permutation of the practiced interval. Hence, $A[0 \ldots n_d+\epsilon'-1]$ can be though of as a short term memory where the encoded information of the practiced interval is stored. On the other hand, $n_c-\epsilon'$ idle integers and $n_d'$ integers out of the practiced interval are distributed disorderly together at $A[n_d+\epsilon' \ldots n-1]$. 

In retrieval phase, the stored information is retrieved from the short term memory $A[0 \ldots n_d+\epsilon'-1]$ to construct the sorted permutation of the practiced interval. The short term memory encodes $n_d+n_c$ integers with $n_d+\epsilon'$ permanent records. It is important to note that, if the number of occurrences of a particular integer is $n_i$, then there are $n_i-1$ idle integers in the array. But the node itself represents the integer that is mapped into the ILS through itself. Hence, it is immediate from this definition that the nodes in the short term memory $A[0 \ldots n_d+\epsilon'-1]$ can be processed from right to left backwards and the integers practiced by each node can be expanded over $A[0 \ldots n_d+n_c-1]$ sequentially right to left backwards without collision. At this point, we have two options: sequential or recursive version. But before proceeding, the retrieval phase will be introduced;

\begin{enumerate}[label=\bf{Algorithm \Alph{*}.}, ref=Algorithm \Alph{*}, leftmargin=0pt, itemindent=*, start=3] 
\item \label{algorithm:es_fgd_mul} Retrieve the encoded information from $n_d+\epsilon'$ records of the short term memory $A[0 \ldots n_d+\epsilon'-1]$ to construct sorted permutation of $n_d+n_c$ integers of the practiced interval. Process the records from right to left backwards and expand the integers over $A[0 \ldots n_d+n_c-1]$ sequentially right to left backwards.
\end{enumerate}

\begin{enumerate}[label=\bf{C\arabic{*}.}, ref=C\arabic{*}, itemindent=* , nosep]
\item initialize $i = n_d+\epsilon'-1$ and $p = n_d + n_c -1$;
\item check MSB of $A[i]$;\label{algo19:item0}
\begin{enumerate}[label=(\roman{*}), ref=(\roman{*}), itemindent=* , nosep]
\item if MSB of $A[i]$ is $1$, then it is a node. Hence, decode from the record of the node the number of idle integers practiced by the node to $k$ (does not include the integer that create the node) and absolute position of the node to $j$ and decrease $i$ by one;\label{algo19:item1}
\item otherwise, $A[i]$ is an idle integer brought immediately after a node. Hence, get the absolute position of the node from the idle integer to $j$ and get the number of idle integers practiced by the node from its record at $A[i-1]$ to $k$ (does not include the integer that create the node) and decrease $i$ by two;\label{algo19:item2}
\end{enumerate}
\item retrieve the integer from the ILS: absolute position $j$ of the node cues the recall of the integer using the inverse hash function. Then copy the integer to $A[p-k \ldots p]$, decrease $p$ by $k+1$ and goto step \ref{algo19:item0};\label{algo19:item3}
\end{enumerate}

\begin{description}[leftmargin = 0pt]
\item [{\bf Sequential Version}] After storing the encoded information into the short term memory, $n_c-\epsilon'$ idle integers and $n_d'$ integers out of the practiced interval are distributed disorderly together at $A[n_d+\epsilon' \ldots n-1]$. If we partition $A[n_d+\epsilon' \ldots n-1]$ selecting the pivot equal to $\delta$, then idle integers are clustered after the short term memory. Therefore, \ref{algorithm:es_fgd_mul} can immediately be used to retrieve. Hence, the structure of the sequential version becomes;
\begin{enumerate}[label=\bf{Algorithm \Alph{*}.}, ref=Algorithm \Alph{*}, leftmargin=0pt, itemindent=*, start=4] 
\item \label{algorithm:es_fgd_iter_mul} In each iteration, construct sorted permutation of $n_d+n_c$ integers of the practiced interval at the beginning of the array.
\end{enumerate}
\begin{enumerate}[label=\bf{D\arabic{*}.}, ref=D\arabic{*}, itemindent=* , nosep]
\item find $\min(A)$ and $\max(A)$;\label{algo20:item1}
\item initialize $\epsilon$ using Eqn.~\ref{eqn:epsilon}, $\delta = \min(A)$, $\delta' = \max(A)$ and reset counters;\label{algo20:item2}
\item practice all the integers in the interval $[\delta,\delta+n-\epsilon-1]$ using \ref{algorithm:es_fgl_mul};\label{algo20:item3}
\item store encoded information of the practiced interval using \ref{algorithm:es_fgp_mul};\label{algo20:item4}
\item in-place partition $A[n_d+\epsilon' \ldots n-1]$ clustering $n_c - \epsilon'$ idle integers at the beginning;\label{algo20:item5}
\item retrieve the sorted permutation of the practiced interval using \ref{algorithm:es_fgd_mul};\label{algo20:item6}
\item if $n_d'=0$ exit. Otherwise set $A=A[n_d+n_c \ldots n-1]$, $n = n_d'$, $\delta = \delta'$, $\delta' = \max(A)$, reset counters, calculate $\epsilon$ using Eqn.~\ref{eqn:epsilon} and goto step \ref{algo20:item3}.\label{algo20:item7}
\end{enumerate}

\begin{rem}
$\min(A)$ and $\max(A)$ need not be found in step \ref{algo20:item1}. Instead, if $\delta = 0$ and $\delta' = \max(\mathbb{U})$ the algorithm sorts the integers in the range $[0,n-\epsilon-1]$ during the first iteration (or recursion). However, if there is not any integer in this interval, \ref{algorithm:es_fgl_mul} finds $\delta'=\min(A)$ in step \ref{algo20:item3} in $\mathcal{O}(n)$ time, and continues with the integers in $[\delta', \delta'+n-\epsilon-1]$.
\end{rem}

\begin{rem}
Sequential version of associative sort technique is on-line in the sense that after each step \ref{algo20:item6}, $n_d+n_c$ integers are added to the sorted permutation at the beginning of the array and ready to be used.
\end{rem}

\item [{\bf A Different Approach.}] Instead of using Eqn.~\ref{eqn:epsilon} to calculate the maximum value of $\epsilon$, there is another approach to solve the same problem possibly more efficiently;
\begin{enumerate}[label=(\roman{*}), itemindent=* , nosep]
\item practice all the integers in the interval $[\delta,\delta+n-1]$ by mapping them into $Im[0\ldots n-1]$ over $A[0\ldots n-1]$. However, during practicing, if the number of idle integers practiced by a particular node reaches exactly $2^{w-1-\log n}$, then increase $\epsilon'$ which counts the exact number of nodes that has practiced at least $2^{w-1-\log n}$ idle integers;\label{algo21:item1}
\item retrieve the integers in $A[n-\epsilon' \ldots n-1]$ back to the array space; \label{algo21:item2}
\item shift subspace to the right by $\epsilon'$;\label{algo21:item3}
\item store encoded information of the practiced interval  using \ref{algorithm:es_fgp_mul};\label{algo21:item4}
\item partition $A[n_d+\epsilon' \ldots n-1]$ clustering $n_c - \epsilon'$ idle integers to the beginning;\label{algo21:item5}
\item retrieve the sorted permutation of the practiced interval using \ref{algorithm:es_fgd_mul};\label{algo21:item6}
\item if $n_d'=0$ exit. Otherwise set $A=A[n_d+n_c \ldots n-1]$, $n = n_d'$, $\delta = \delta'$, $\delta' = \max(A)$, reset counters and goto step \ref{algo21:item1}.\label{algo21:item7}
\end{enumerate}
In this case, instead of using the maximum value of $\epsilon$, its exact value $\epsilon'$ is counted and used which may improve the overall efficiency.

\item [{\bf Recursive Version}] Saving $n_d$, $\epsilon'$ and $\delta$ in stack space, we can recursively call \ref{algorithm:es_fgl_mul} and \ref{algorithm:es_fgp_mul}. Although the exact number of integers to be sorted in the next level of recursion is $n_d'$, the overall number of integers in that recursion is $n=n_d'+n_c-\epsilon'$ where $n_c-\epsilon'$ of them are idle integers of the previous recursion and meaningless. However, these idle integers increase the interval of range of integers spanned by the ILS improving the overall time complexity in each level of recursion. The recursion can continue until no any integer exists. In the last recursion, retrieval phase can begin to construct the sorted permutation of $n_d+n_c$ integers from $n_d+\epsilon'$ records stored in the short term memory $S[0\ldots n_d+\epsilon'-1]$ of that recursion and expand over $S[0 \ldots n-1]$ right to left backwards. Each level of recursion should return the total number of integers copied on the array to the higher level to let it know where it will start to expand its interval. It should be noticed that, in the recursive version of the technique, there is no need to partition $n_c-\epsilon'$ idle integers from $n_d'$ unpracticed integers. Hence, one step is canceled improving the overall efficiency.

\end{description}

\begin{description}[leftmargin = 0pt]

\item [{\bf Complexity}] of the algorithm depends on the range and the number of integers. In each iteration (or recursion) the algorithm is capable of sorting integers that satisfy $A[i] - \delta + \epsilon < n$ where $\epsilon$ is defined by Eqn.~\ref{eqn:epsilon} and upper bounded by $n/2$. Hence, at worst case ($n= 2^{w-1}$), the integers that satisfy $A[i] - \delta < n/2$ are sorted in the first pass. This means that, given uniformly distributed $n=2^{w-1}$ integers $A[0 \ldots n-1]$ each in the range $[0,n-1]$, the complexity is the recursion $T(n) = T(\frac{n}{2}) + \mathcal{O}(n)$ yielding $T(n) = \mathcal{O}(n)$.

\item[{\bf Best Case Complexity.}] Given $n$ integers $A[0 \ldots n-1]$, if $n-1$ integers satisfy $A[i] - \delta < n/2$, then these are sorted in $\mathcal{O}(n)$ time. In the next step, there is one integer left which implies sorting is finished. As a result, time complexity of the algorithm is lower bounded by $\Omega(n)$ in the best case.

\item [{\bf Worst Case Complexity.}] Given $n$ integers $A[0 \ldots n-1]$ and $m=\beta n$, if there is only $1$ integer available in practiced interval at each iteration (or recursion) until the last, in any $j$th step, the only integer $s$ that will be sorted satisfies $s < \frac{jn-(j-1)}{2}$ which implies that the last alone integer satisfies $s < \frac{jn-(j-1)}{2} \le \beta n$ from where we can calculate $j$ by $j \le \frac{2\beta n-1 }{n-1}$.
In this case, the time complexity of the algorithm is,
\begin{equation}
\mathcal{O}(n) + \mathcal{O}(n-1) + \dotsc + \mathcal{O}(n-j) = (j+1) \mathcal{O}(n) -\mathcal{O}(j^2) < (2\beta+1) \mathcal{O}(n)
\end{equation}

Therefore, the algorithm is upper bonded by $(2\beta+1) \mathcal{O}(n) = \mathcal{O}(2m+n)$ in worst case.

\item[{\bf Average Case Complexity.}] Given $n$ integers $A[0 \ldots n-1]$, if $m = \beta n$ and the integers are uniformly distributed, this means that $\frac{n}{2\beta}$ integers satisfy $A[i] < \frac{n}{2}$. Therefore, the algorithm is capable of sorting $ \frac{n}{2\beta}$ integers in $\mathcal{O}(n)$ time during first pass. This will continue until all the integers are sorted. The sum of sorted integers in each iteration can be represented with the series,
\begin{equation} \label{eqn:ud_1}
\frac{n}{2\beta}+\frac{n(2\beta-1)}{4\beta^2}+\frac{n(2\beta-1)^2}{8\beta^3}+\dotsc+\frac{n(2\beta-1)^{k-1}}{2^k\beta^{k}}+\dotsc
\end{equation}

It is reasonable to think that the sorting ends when one term is left which means the sum of $k$ terms of this series is equal to $n-1$, from where we can calculate the number of iteration or dept of recursion $k$ which is valid when $\beta > \frac{1}{2}$,
\begin{equation} \label{eqn:ud_4}
\frac{1}{n} = \frac{(2\beta-1)^{k-1}}{(2\beta)^{k}}
\end{equation}
It is seen from Eqn.~\ref{eqn:ud_4} that when $m = n$, i.e., $\beta=1$, number of iteration or dept of recursion becomes $k=\log{n}$. It is known that each step takes $\mathcal{O}(n)$ time. Therefore, the time complexity of the algorithm is,
\begin{equation}\label{eqn:ud_5}
\mathcal{O}(n) \bigl( \frac{(2\beta-1)}{2\beta} + \frac{(2\beta-1)^2}{(2\beta)^2} +\dotsc+ \frac{(2\beta-1)^{k-1}}{(2\beta)^{k-1}} \bigr)
\end{equation}
from where we can obtain by defining $x= \frac{(2\beta-1)}{2\beta}$,
\begin{equation}\label{eqn:ud_6}
\mathcal{O}(n) \bigl( 1 +  x + x^2 + x^3 + \cdots + x^{k-1} \bigr) = \mathcal{O}(n) (\frac{1}{1-x} - \frac{x^{k-1}}{1-x}) < 2\beta \mathcal{O}(n)
\end{equation}
which means that the algorithm is upper bounded by $ 2\beta \mathcal{O}(n)$ or $2\mathcal{O}(m)$ in the average case.

\end{description}

\section{Conclusions}
\label{chap:summaryandconclusion}

In this study, in-place associative integer sorting technique is introduced. Using the technique, the main difficulties of distributive sorting algorithms are solved by its inherent three basic steps namely (i) {\em practicing}, (ii) {\em storing} and (iii) {\em retrieval} which are three main stages in the formation and retrieval of memory in cognitive neuroscience. The technique is very simple and straightforward and around 30 lines of C code is enough.

The technique sorts the integers using $\mathcal{O}(1)$ extra space in $\mathcal{O}(n+m)$ time for the worst, $\mathcal{O}(m)$ time for the average (uniformly distributed integers) and $\mathcal{O}(n)$ time for the best case. It shows similar characteristics with bucket sort and distribution counting sort and hence can be thought of as {\em in-place associative bucket sort} or {\em in-place associative distribution counting sort}. However, it is time-space efficient than both. The ratio $\frac{m}{n}$ defines the efficiency (time-space trade-offs) letting very large arrays to be sorted in-place. Furthermore, the dependency of the efficiency on the distribution of the integers is $\mathcal{O}(n)$ which means it replaces all the methods based on address calculation, that are known to be very efficient when the integers have known (usually uniform) distribution and require additional space more or less proportional to $n$. Hence, associative sort asymptotically outperforms all content based sorting algorithms when $\frac{m}{n}=c$ and $c$ is the efficiency constant determined by the other sorting algorithms regardless of how large is the array.

The technique seems to be very flexible, efficient and applicable for other problems, as well, such as membership and range queries, hashing, searching, element distinction, succinct data structures, gaining space, etc. For instance, gaining space is an inherent step of associative sort which improves its performance and can be used explicitly, as well.

The drawbacks of the algorithm is that it is unstable as well as suitable for value-sorting. But, an ILS can create other subspaces and associations using the idle integers that were already practiced by manipulating either their position or value or both. Hence, different techniques can be developed to solve such problems.


\end{document}